\documentclass[prl,twocolumn,showpacs,groupedaddress]{revtex4}

\usepackage{graphicx}

\begin{document}

\title{\bf Multi-Welled Tunneling Model for the Magnetic-Field Effect in 
Ultracold Glasses }
 
\author{ Giancarlo Jug \cite{P} } 

\affiliation{ 
Dipartimento di Fisica e Matematica, Universit\`a dell'Insubria, Via
Valleggio 11, 22100 Como, Italy \\
CNISM -- Unit\`a di Ricerca di Como and INFN -- Sezione di Pavia, Italy }   

\date{\today }

\begin{abstract}
Puzzling observations of both thermal and dielectric responses in 
multi-silicate glasses at low temperatures $T$ to static magnetic fields 
$B$ have been reported in the last decade and call for an extension of 
the standard two-level systems tunneling model. An explanation is proposed, 
capable of capturing at the same time the $T$- and $B$-dependence of the 
specific heat $C_p$ and of the dielectric constant $\epsilon$ in these 
glasses. This theory points to the existence of anomalous multi-welled 
tunneling systems in the glasses -- alongside the standard two-level 
systems  -- and indications are given for glasses which should achieve 
larger electric magnetocapacitive enhancements.
\end{abstract}


\maketitle

The last decade has seen much renewed interest in the physics of cold 
non-metallic glasses, materials displaying some universal physical 
properties attributed to the low-energy excitations characterising most 
amorphous solids. The two-level systems (2LS) tunneling model (TM) 
\cite{ref1} has been rather successful in explaining a variety of 
interesting phenomena dealing with the thermal, dielectric and acoustic 
properties of structural glasses at temperatures $T<$ 1 K. Limitations 
and failures of the 2LS TM (treating cooperative motion in terms of 
single particles, mostly) on the other hand, have been also discussed 
\cite{ref2}.

These materials do not present, normally, any remarkable magnetic-field 
response phenomena other than possibly a weak contribution from trace 
paramagnetic Fe-impurities. It came thus as a great surprise when 
measurements showed \cite{ref3} that in some multi-component silicate 
glasses (but not in pure {\it a}-SiO$_2$) one is able to observe changes 
in the dielectric constant $\epsilon(T,B)$ and already in magnetic fields 
$B$ as weak as a few Oe. A typical glass showing a strong response has some 
100 ppm Fe$^{3+}$ and the composition Al$_2$O$_3$-BaO-SiO$_2$ (in short 
AlBaSi-O, in this paper). Measurements made on a thick sol-gel fabricated 
AlBaSi-O film showed changes in 
$\delta\epsilon/\epsilon=[\epsilon(B)-\epsilon(0)]/\epsilon(0)$ of
order 10$^{-4}$ and characterised by an {\it enhancement} peaking around 
0.03 T for 10 mK $< T <$ 200 mK then followed by a {\it reduction} of 
$\epsilon$ for $B>$ 0.1 T. A further enhancement was also observed 
at much higher fields ($B>$ 10 T). 

Another, cleaner, multi-component silicate glass (borosilicate BK7, with 
6 ppm Fe$^{3+}$) and a dirtier one (borosilicate Duran, with 120 ppm
Fe$^{3+}$) have shown similar -- but weaker -- magnetic anomalies, 
$\vert\delta\epsilon/\epsilon\vert\sim$ 10$^{-5}$, seemingly excluding 
the paramagnetic impurities as their source \cite{ref4}. Yet another 
multi-silicate glass, {\it a}-SiO$_{2+x}$C$_y$H$_z$ was investigated 
\cite{ref5}, confirming the unusual findings in AlBaSi-O. A convincing 
explanation for the unusual electric magnetocapacitance behavior of these 
cold glasses has not yet been found.     

Recently, some consensus has been gained by the idea of a coupling of the 
standard 2LS to the magnetic field via nuclei in the glasses carrying an 
electric quadrupole moment as well as a magnetic dipole one \cite{ref6}. 
The nuclear mechanism is supported by some features of the observed 
magnetic-field dependence of the polarization echo (PE) experiments in the 
mentioned multi-silicate glasses in the millikelvin range \cite{ref7}. 
Moreover the amplitude of the PE in Glycerol glass was shown to become 
strongly $B$-dependent only upon deuteration and thus the introduction of 
quadrupole-moment carrying nuclei in the glass \cite{ref8}. However, though 
pure {\it a}-SiO$_2$ (devoid of quadrupole-moment carrying nuclei) shows no 
PE-amplitude magnetic field dependence \cite{ref7}, Glycerol glass -- 
deuterated or not -- shows no measurable $B$-dependence in its dielectric 
constant \cite{ref9}. The nuclear approach is also unable to account for the 
magnitude and features of the $B$- and $T$-dependence of 
$\delta\epsilon/\epsilon$ for the multi-silicate glasses \cite{ref10} and 
the (also unusual) $B$- and $T$-dependence of the heat capacity $C_p$ of the 
latter -- not entirely linked to their Fe impurity contents 
\cite{ref11,ref12} -- is not even addressed. To add to the mystery, the 
acoustic response (also linked to the 2LS coupling to phonons) of 
borosilicate glasses BK7 and AF45 has been found to be independent of $B$ 
\cite{ref13}. Table I summarises this rather puzzling experimental situation. 
Therefore, either the nuclear explanation is specific to the PE magnetic 
effect, or an entirely new explanation should be found for all of the 
observations.

\begin{table}[ht]
\begin{tabular}{|l|c|cccc|}
\hline
glass type & Ref. & $\delta C_p$ & $\delta\epsilon$ & $\delta v_s$ &
$\delta A_{PE}$ \\
\hline \hline
{\it a}-SiO$_2$ & \cite{ref11},\cite{ref4},\cite{ref7} & NO & NO & ? & NO \\
\hline
{\it a}-SiO$_{2+x}$C$_y$H$_z$ & \cite{ref5} & ? & YES & ? & ? \\
\hline
AlBaSi-O & \cite{ref11},\cite{ref3},\cite{ref7} & YES & YES & ? & YES \\
\hline
Duran & \cite{ref11},\cite{ref4},\cite{ref7} & YES & weak & ? & YES \\
\hline
BK7 & \cite{ref11},\cite{ref4},\cite{ref13},\cite{ref7}
& weak & weak & NO & YES \\
\hline
AF45 & \cite{ref13} & ? & ? & NO & ? \\
\hline \hline
Glycerol & \cite{ref9},\cite{ref8} & ? & NO & ? & weak \\
\hline
d-Glycerol & \cite{ref9},\cite{ref8} & ? & NO & ? & YES \\
\hline
\end{tabular}
\caption[1]{Presence of magnetic-field induced variations in the physical
properties of some cold glasses. $C_p$: heat capacity, $\epsilon$: 
dielectric constant, $v_s$: sound velocity, $A_{PE}$: PE amplitude 
[?: no investigation known]}
\label{tabl1}
\end{table}

The purpose of this Letter is to begin to give a rationale to the 
situation in Table I, as well as to stimulate further experimental and 
theoretical research. A novel explanation, already shown to
account for the unusual behavior of $C_p(T,B)$ in the multi-silicates 
\cite{ref12}, is here shown to explain the behavior of the dielectric 
constant as well. This simple theory is centered on the observation, in 
computer simulations and experiments \cite{ref14}, that multi-silicate 
glasses present in their atomic structure both the connected network of 
SiO$_4$-tetrahedra and a collection of pockets and channels of 
non-networking ions showing a tendency to form microaggregates and to 
partially destroy the SiO$_4$-network. Fig. 1 shows as an example a snapshot 
of a simulation of the (Na$_2$O)-3(SiO$_2$) glass illustrating such 
situation. The present theory proposes that the magnetic effects arise from 
{\it anomalous} tunneling systems (ATS) forming in the cooling of such 
structure within the non-networking (or network-modifying, NM) pockets and 
channels, whilst the SiO$_4$-network remains the nest of the ordinary 
non-magnetic 2LS. 

\begin{figure}[ht]
\includegraphics[width=0.40\textwidth]{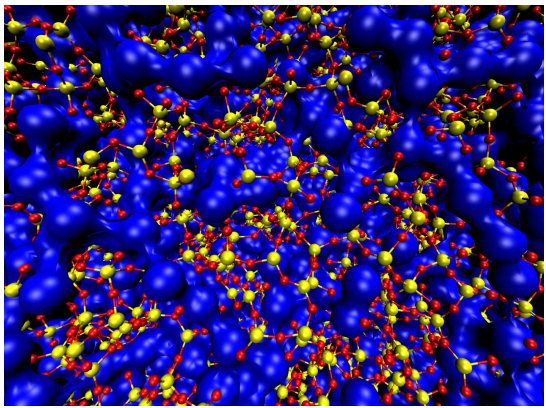} \vskip0mm
\caption[2]{ Molecular dynamics snapshot of the structure of sodium
trisilicate at 2100 K at the density $\rho=2.2$ g cm$^{-3}$. The blue
spheres that are connected to each other represent the (NM) Na atoms. The 
Si-O network is drawn by (NF) yellow (Si) and red (O) spheres that are 
connected to each other by covalent bonds (from \cite{ref14}, by permission).
 }
\label{fig1}
\end{figure}

In order to couple the ATS to the magnetic field, a simple 3D generalization 
of the 2LS TM is in order. As is known \cite{ref1}, in the TM the cold 
glass is thought to have few remaining degrees of freedom rappresented by 
{\it fictitious} ``particles'', each moving quantum-mechanically within a 1D 
double-welled potential. At low temperatures only the ground states of the 
individual wells $\vert i \rangle$ ($i$=1, 2) are relevant and in this 
representation the Hamiltonian of a single 2LS reads 
$H_0=-\frac{1}{2}\Delta\sigma_z-\frac{1}{2}\Delta_0\sigma_x$ ($\sigma_{\mu}$ 
Pauli matrices) with $\Delta$ the ground-state energy asymmetry between the 
two wells and $\Delta_0$ the barrier's transparency. These two parameters 
are linked to the real potential's details as discussed in the TM literature 
and are taken, normally, to be distributed in the glass so that $\Delta$ and 
$\ln\Delta_0$ (roughly, the potential barrier) have a uniform distribution:
${\cal P}_{2LS}(\Delta,\Delta_0)=\bar{P}/\Delta_0$, $\bar{P}$ being a 
material-dependent constant. This description holds for the network-forming 
(NF) TS. For the NM ATS instead, the simplest 3D generalization of the 2LS 
TM is that of other {\it fictitious} charged particles, each moving in a 
multi-welled 3D potential (see Fig. 2) and coupling to the magnetic field 
through their orbital motion. For the simplest case of $n_w=3$ potential 
wells, one can use:
\begin{equation}
H_0 = \left ( \begin{array}{ccc}
E_1 & D_0 e^{i\phi/3} & D_0 e^{-i\phi/3} \\
D_0 e^{-i\phi/3} & E_2 & D_0 e^{i\phi/3} \\
D_0 e^{i\phi/3} & D_0 e^{-i\phi/3} & E_3
\end{array} \right )
\label{hamiltmx}
\end{equation}
where $D_0\simeq \hbar\Omega e^{-U_B/\hbar\Omega}$ is some 3D barrier's 
transparency ($U_B$ barrier height and $\Omega$ single-well frequency),
$E_1, E_2, E_3$ are the single wells' ground-state energy asymmetries 
($E_1+E_2+E_3=0$, say) and 
\begin{eqnarray}
\phi&=&2\pi\Phi({\bf B})/\Phi_0 \\
\Phi({\bf B})&=&{\bf B}\cdot{\bf S}_{\triangle}=BS_{\triangle}\cos\beta
\nonumber
\label{abphase}
\end{eqnarray}
is some Aharonov-Bohm phase for a ``particle'' carrying charge $q$ tracing 
a closed path of area $S_{\triangle}$ threaded by a magnetic flux 
$\Phi({\bf B})$ ($\Phi_0\equiv h/|q|=\varphi_0\vert e/q\vert$ being the 
appropriate flux quantum ($\varphi_0$ being the elementary one)).     

\begin{figure}[ht]
\includegraphics[angle=0, width=0.33\textwidth]{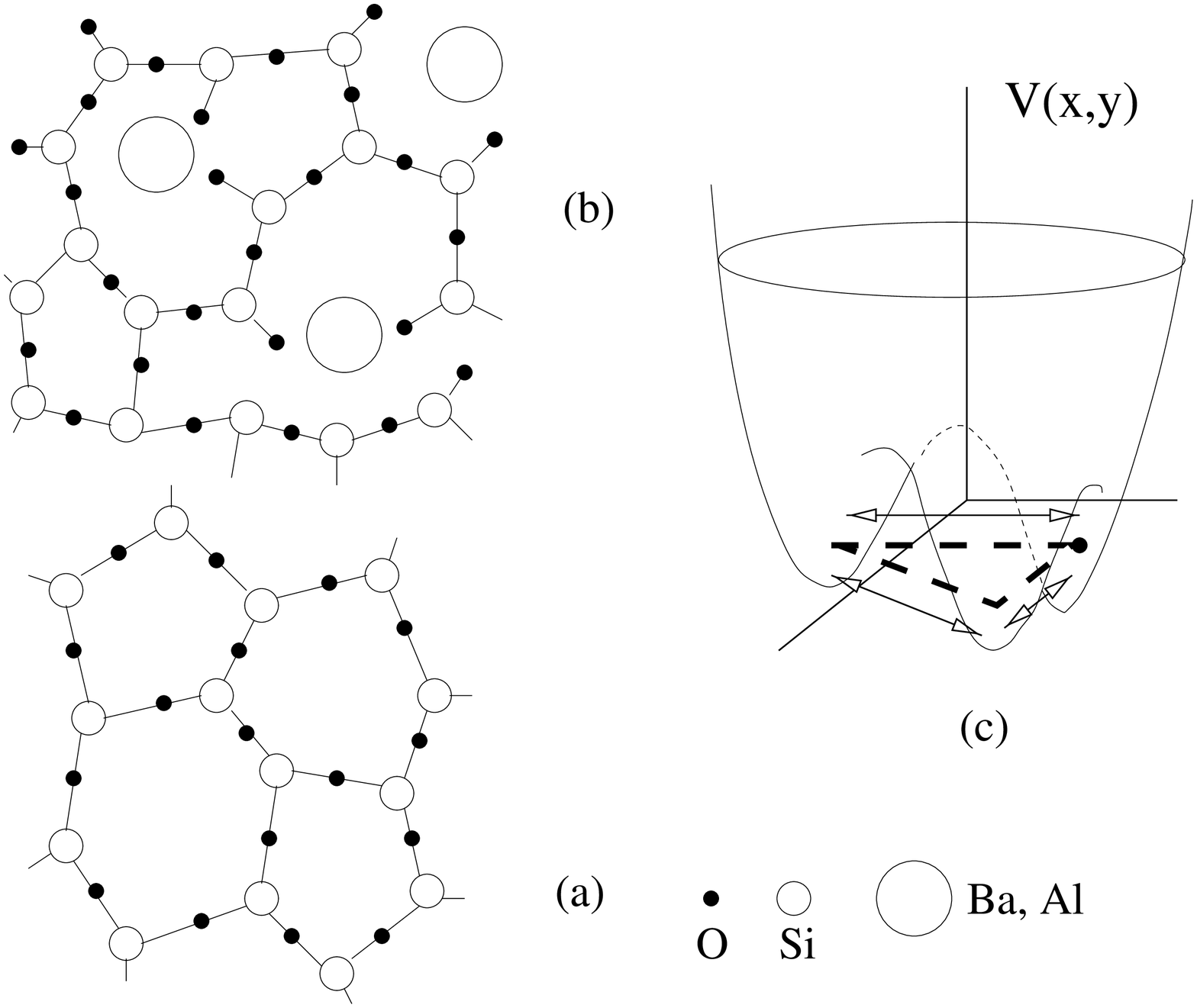}\vskip0mm
\caption[2]{ 2D representation of the plausible source of magnetic-field 
sensitive (anomalous) tunneling systems in (e.g.) the AlBaSi-O glass. The 
tight vitreous-SiO$_2$ structure (a) is broken up by the Al- and (large) 
Ba-atoms (b), thus leaving many metal ions free to move as ``particles'' in 
a 3D tunneling potential characterised by $n_w$-minima, with $n_w\ge3$ (c). }
\label{fig2}
\end{figure}

In this model the choice $D_0>0$ can be thought of as arising from the 
coherent tunneling motion of a small cluster of NM-ions; this will lead, 
as is seen, to high values of the product $|q|S_{\triangle}D_0$. For 
$D\equiv\sqrt{E_1^2+E_2^2+E_3^2}\ll D_0$ and weak fields ($\phi\ll 1$) the 
lowest energy gap of the model is seen to open with increasing magnetic
field according to the simplified expression \cite{ref12} 
$\Delta {\cal E}\simeq\sqrt{D_0^2\phi^2+D^2}$ containing the main physics 
of this model (regardless of the value of $n_w$). One more assumption of 
this theory (seen to explain the nearly-flat $T$-dependence of $C_p$ for 
$B=0$ in some temperature range for these glasses \cite{ref12}) is to take 
the parameters' distribution uniform for $U_B$, but favoring near-degeneracy 
(to a degree fixed by a lower bound $D_{min}\not=0$ for $D$) for the 
$\{E_i\}$:
\begin{equation}
{\cal P}_{ATS}(\{E_i\},\cdots;D_0)=\frac{P^*}{(E_1^2+E_2^2+E_3^2
+\cdots)D_0}.
\label{3lsdistr}
\end{equation}
This {\it anomalous} distribution for the ATS can be thought of as arising 
from a degree of {\it devitrification} in the material, as measured by the 
parameter $P^*$. In fact, it is reported \cite{ref15} that 
thick glass films prepared with the sol-gel technique (such as the AlBaSi-O 
films of the experiments) are multiphase materials with microcrystals 
embedded within an amorphous glassy matrix \cite{ref16,ref17}. Indeed, 
amorphous solids with the general composition Al$_2$O$_3$-MgO-CaO-SiO$_2$
are termed ``glass ceramics'' in the literature \cite{ref16} owing to 
partial devitrification occurring. It seems thus reasonable to imagine 
that in these glasses some NM-ions provide nucleation centres for the 
microcrystals and that, therefore, TS presenting near-symmetric wells will 
be favoured in the neighborhood of and inside such crystallites.

This approach has provided a good description \cite{ref12} of the $C_p(T,B)$ 
data for AlBaSi-O and Duran \cite{ref11}; one can treat the ATS as 
{\it effective} 2LS having gap $\Delta{\cal E}$ for ``weak'' fields. Within 
this picture, the linear-response quasi-static resonant contribution to the 
polarizability is
\begin{equation}
\alpha^{RES}_{\mu\nu}=\int_0^{\infty}\frac{dE}{2E}
{\cal G}_{\mu\nu}\left (\{\frac{E_i}{E}\};{{\bf p}_i}\right )
\tanh(\frac{E}{2k_BT})\delta(E-\Delta{\cal E})
\label{respolariz}
\end{equation}
where
\begin{equation}
{\cal G}_{\mu\nu}\left (\{\frac{E_i}{E}\};{{\bf p}_i}\right )=
\sum_{i=1}^{n_w}p_{i\mu}p_{i\nu}-\sum_{i,j}\frac{E_iE_j}{E^2}p_{i\mu}p_{j\nu}
\label{geomcorr}
\end{equation}
contains the single-well dipoles ${\bf p}_i=q{\bf a}_i$. This
expression assumes vanishing electric fields and no TS-TS interactions, a 
situation which does not wholly apply to the experiments. To keep the theory 
simple one can still use Eq. (\ref{respolariz}) and the analogous one for 
the relaxational contribution to the polarizability. Eq. (\ref{respolariz}) 
must be averaged over the random energies' distribution (\ref{3lsdistr}) 
($[\dots]_{av}$, responsible for the high sensitivity to weak fields) and 
over the dipoles' orientations and strengths ($\overline{(\dots)}$). For a 
collection of ATS with $n_w>2$ this averaging presents serious difficulties 
and one must resort to the decoupling:
\begin{equation}
\overline{{\cal G}_{\mu\nu}\delta(E-\Delta{\cal E})}\simeq
\overline{{\cal G}_{\mu\nu}}\cdot\overline{\delta(E-\Delta{\cal E})},
\label{decoupl}
\end{equation}
where $\overline{[\delta(E-\Delta{\cal E})]_{av}}=g_{ATS}(E,B)$ is the 
fully-averaged density of states. To calculate $\overline{{\cal G}_{\mu\nu}}$, 
one can envisage a fully isotropic distribution of planar $n_w$-polygons to 
obtain:
\begin{equation}
\overline{{\cal G}_{\mu\nu}}=\frac{1}{3}\left ( \frac{n_w}{n_w-1} \right )
\overline{p_i^2}\frac{(n_w-2)E^2+D_0^2\phi^2}{E^2}\delta_{\mu\nu}.
\label{isoglass}
\end{equation}
The second term in the numerator of Eq. (\ref{isoglass}) gives rise to a
peak in $\delta\epsilon/\epsilon$ at very low $B$, while the first term
(present only if $n_w>2$) gives rise to a {\it negative} contribution to
$\delta\epsilon/\epsilon$ at larger $B$ which can win over the enhancement
term for all values of $B$ if $D_{0max}\gg D_{0min}$ ($D_{0min}$, $D_{0max}$ 
corresponding to cutoffs in the distribution of ATS energy barriers $U_B$). 
The observations in Duran and BK7 indeed show a significant depression of 
$\epsilon(B)$ for weak fields \cite{ref4}, thus giving direct evidence for 
the existence of ATS with $n_w>2$ in the multisilicate glasses. Carrying out 
the averaging $[\dots]_{av}$ one gets analytical expressions for the 
polarizability; the uniform average over orientation angles $\beta$ must be 
performed numerically.

\begin{figure}[h]
\includegraphics[width=0.48\textwidth]{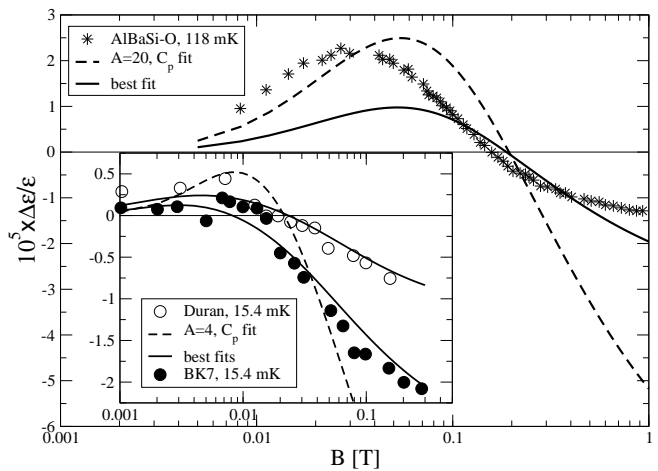} \vskip0mm
\caption[2]{ Dielectric constant's change (real part) in a magnetic field, 
measured and from the present theory, for the multicomponent AlBaSi-O glass 
at $T=$ 118 mK (driving field 15 kV m$^{-1}$, driving frequency 1 kHz) 
[inset: data and theoretical calculations for the BK7 and Duran glasses at 
15.4 mK]. Data from Ref. \cite{ref4}. }
\label{fig3}
\end{figure}

Fig. 3 shows the change due to a magnetic field in the dielectric constant
$\epsilon(T,B)=\epsilon_0+\alpha(T,B)$ for the AlBaSi-O glass within the 
present oversimplified theory, which is used to fit the data at 118 mK
\cite{ref4}. It is seen that using the parameters extracted from fitting 
this theory \cite{ref12} to the $C_p(T,B)$ data \cite{ref11} and a value 
$A=$ 20.0 $\times$ 10$^{-5}$ for the dimensionless prefactor   
$A\equiv\overline{p_i^2}P^*/(\epsilon_0\epsilon_rD_{min})$
one can reproduce all the qualitative features of the puzzling 
$B$-dependence of the magnetocapacitance for this glass. A best fit to the 
118 mK experimental data yields the parameters reported in Table II. The 
$T$-dependence of the calculated magnetocapacitance is also in qualitative 
agreement with the experimental data \cite{ref3,ref4}. In a comparison with 
data for the BK7 and Duran glasses \cite{ref4} (at the much lower only 
available temperature of 15.4 mK, however) the present theory reproduces 
the main features correctly. The values of $D_{0min}|q/e|S_{\triangle}$ and 
$D_{0max}|q/e|S_{\triangle}$ are very similar to those used to fit the 
$C_p(T,B)$ data \cite{ref12}; the small values of $D_{0min}/D_{0max}$ and 
of $A$ for BK7 denote a much reduced presence of microcrystallites in the 
material. The values of $D_{min}$ are lower (due to the strong electric 
field applied) and more realistic than those used in the analysis of 
$C_p(T,B)$, in all cases confirming the consistency of the assumption 
$\vert E_i\vert/D_0\ll1$ in this theory.

\begin{table}[ht]
\begin{tabular}{|l|cccc|}
\hline
glass & $A$ & $D_{min}$ &
$D_{0min}\vert q/e \vert S_{\triangle}$ &
$D_{0max}\vert q/e \vert S_{\triangle}$ \\
type  &  & [mK]  &  [K $\AA^2$]  &  [K $\AA^2$] \\
\hline \hline
AlBaSi-O & 1.42$\times$10$^{-4}$ & 30 & 9.72$\times$10$^4$
& 1.69$\times$10$^5$ \\
\hline
Duran & 0.98$\times$10$^{-5}$ & 34 & 1.83$\times$10$^4$
& 6.28$\times$10$^5$  \\
BK7 & 0.95$\times$10$^{-5}$ & 24 & 4.16$\times$10$^4$
& 1.09$\times$10$^6$ \\
\hline
\end{tabular}
\caption[1]{Local ATS parameters extracted from this fit}
\label{tabl2}
\end{table}

Finally, the present simplified theory has focused on the weak $B$-field 
regime (up to 1 T). Given the large values of $D_0|q/e|S_{\triangle}$ 
thus extracted (indicating that small coherent clusters of some 5 to 10 
NM-ions are involved in the magnetic-sensitive tunneling \cite{ref12}) one 
can expect the low-$B$ expression employed for the gap to break down around
some larger field 
$B^*\sim\varphi_0(D_{0min}/D_{0max})/(2\pi\vert q/e \vert S_{\triangle})$ 
above which the gap grows sub-linearly with $\phi(B)$. As will be shown 
elsewhere, this is in turn responsible for the second enhancement of 
$\delta\epsilon/\epsilon$ observed in the experiments for $B>B^*$ ($B^*$ 
being rather material-dependent).  

In summary, the ingredients of the present two-species TM together with the 
reasonable assumption of partial devitrification in the films (which can
be checked through X-ray analysis) allows for a first good understanding 
of the puzzle of the magnetocapacitance in the cold multi-silicate glasses. 
The absence of a magnetoacoustic response in such glasses with ATS can be
understood in terms of the much higher resonance frequencies of the 
NM-pockets and channels, and experiments should be done in such conditions. 
As for the PE-experiments, the orbital-coupling approach with the inclusion of 
TS-TS interactions has been shown to provide a partial explanation for the 
$B$-dependence of the PE-amplitude at ultralow temperatures \cite{ref18}.
The present theory, interaction improved, is expected to also provide an
explanation for such data, the so-called isotope effect \cite{ref8} being, 
possibly, fabrication- rather than isotope-related. Further experiments
are needed: clearly, if the magnetic effects are due to quadrupole moments
then the response should scale with the quadrupole-carrying nuclear 
concentration. If tunneling paramagnetic moments are involved, as is
suggested by a localised TM also capable of providing a good explanation for
the $C_p$ and $\epsilon$ data \cite{ref19}, then the magnetic response should 
scale with the Fe-concentration. In the present approach the response scales 
with the NM-ions' concentration and with the degree of devitrification, thus a
larger magnetic response than thus far observed should be found in the best 
ceramic glasses, like for instance Ceran.       

The permission by its Authors to display Fig. 1 is gratefully acknowledged.

\end{document}